\documentclass[10pt,twocolumn]{article}
\usepackage[T1]{fontenc}
\usepackage[utf8]{inputenc}
\usepackage{lmodern}
\usepackage[margin=0.72in,top=0.76in,bottom=0.76in]{geometry}
\usepackage{amsmath,amssymb,booktabs,array,multirow,tabularx}
\usepackage{graphicx,microtype,xurl,enumitem}
\usepackage{tikz}
\usetikzlibrary{arrows.meta,positioning,calc,fit,shapes.geometric}
\usepackage[numbers,sort&compress]{natbib}
\usepackage[colorlinks=true,urlcolor=blue,linkcolor=black,citecolor=blue]{hyperref}
\usepackage[font=small,labelfont=bf]{caption}
\setlist{noitemsep,topsep=2pt}
\newcommand{\AutoTorch}{\textsc{AutoTorch}}
\newcommand{\AutoResearch}{\textsc{AutoResearch}}

\newcommand{\code}[1]{\texttt{#1}}
\definecolor{navy}{RGB}{35,58,85}
\definecolor{softblue}{RGB}{232,241,250}
\definecolor{softgreen}{RGB}{230,242,232}
\definecolor{softgray}{RGB}{243,245,248}
\tikzset{b/.style={draw=navy,rounded corners=2pt,fill=softblue,align=center,inner sep=4pt,font=\scriptsize},
 g/.style={draw=navy,rounded corners=2pt,fill=softgreen,align=center,inner sep=4pt,font=\scriptsize},
 s/.style={draw=navy,rounded corners=2pt,fill=softgray,align=center,inner sep=4pt,font=\scriptsize},
 arr/.style={-Latex,thick,draw=navy},
 dasharr/.style={-Latex,dashed,draw=navy}}
\hypersetup{pdftitle={Agent Stack for ML Training Lifecycle: Generation, Validation, and Optimization},
 pdfauthor={Jaideep Ray, Daniel Tang, Zejun Lin, Haoyue Tang, Lijuan Zhang, Tao Huang, Tommy Li, Aditya Jain, Ankit Goyal, Harsh Jain, Zhuo Zhi, Youmin Han, Santosh Jha, Shuying Liang, Yujie Ai, Jonathan Hung, Ethan Lyu, Abi Liu, Siddharth Singh, Feng Wen, Mahesh Ravindran, Souvik Ray, Yitong Zhou, Animesh Singh}}
\begin{document}
\twocolumn[
\begin{center}
{\LARGE\bfseries Agent Stack for ML Training Lifecycle:\\[3pt] Generation, Validation, and Optimization\par}
\vspace{9pt}
{\small
Jaideep Ray$^{*}$, Daniel Tang$^{*}$, Zejun Lin$^{*}$, Haoyue Tang$^{*}$, Lijuan Zhang$^{*}$, Tao Huang$^{*}$, Tommy Li$^{*}$,\\[2pt]
Aditya Jain$^{*}$, Ankit Goyal$^{*}$, Harsh Jain$^{*}$, Zhuo Zhi$^{*}$, Youmin Han$^{*}$, Santosh Jha$^{*}$, Shuying Liang$^{*}$,\\[2pt]
Yujie Ai$^{*}$, Jonathan Hung$^{*}$, Ethan Lyu$^{*}$, Abi Liu$^{*}$, Siddharth Singh$^{*}$, Feng Wen$^{*}$, Mahesh Ravindran$^{*}$,\\[2pt]
Souvik Ray$^{*}$, Yitong Zhou$^{*}$, Animesh Singh$^{*}$\par}
\vspace{4pt}
{\small LinkedIn\quad\textbullet\quad $^{*}$Equal contribution\quad\textbullet\quad Correspondence: \href{mailto:jaray@linkedin.com}{jaray@linkedin.com}}\par
\end{center}
\vspace{4pt}
\begin{minipage}{0.95\textwidth}
\begin{center}\bfseries Abstract\end{center}
\small
Production model development requires repeated architecture design, platform integration, debugging, training, and evaluation. We present an agent stack deployed at LinkedIn with two independently usable agent loops: AutoTorch for model generation and migration, and AutoResearch for model optimization. AutoTorch generates PyTorch models and training pipelines for TensorFlow migration, XGBoost replacement, and new-model development. Executable validation hooks check platform API compliance, trainability, and conversion fidelity on sampled data, guiding iterative repair before engineer review and full-scale distributed training. AutoResearch explores architecture and training-configuration changes through expert-steered or autonomous experimentation. An external orchestrator manages parallel trials, experiment history, candidate selection, and wall-clock budgets across agent sessions. A shared runtime preserves execution state independently of agent reasoning and allocates GPUs separately from agent execution. Across ranking, content understanding, and recommendation workloads, AutoTorch completed generation and full-scale training in 92 of 115 workflow runs (80\%). Of these, 77 (83.7\%) produced models meeting post-training quality requirements. In a separate cohort, AutoResearch improved the primary task metric on held-out data for 28 of 100 production pipelines. Engineering teams estimate that the stack reduced elapsed end-to-end migration time from approximately eight weeks to two weeks, a $4\times$ speedup.
\end{minipage}
\vspace{14pt}
]
\section{Introduction}
Developing production machine learning models remains a labor-intensive process. Engineers must design model architectures, integrate them with platform-specific training APIs, debug runtime failures, and repeatedly train and evaluate models to achieve target quality. Framework migrations add further complexity, requiring engineers to reproduce existing model behavior while adapting to a new training stack.

\begin{figure*}[t]
\centering
\begin{tikzpicture}[node distance=4mm and 5mm]
\node[s,minimum width=1.5cm,minimum height=1.0cm] (src) {Source model\\TensorFlow / XGBoost};
\node[b,right=of src,minimum width=2.5cm,minimum height=1.0cm] (at) {\textbf{AutoTorch}\\Model generation\\PyTorch + Flyte};
\node[g,right=of at,minimum width=1.55cm,minimum height=1.0cm] (bundle) {Training\\model bundle};
\node[b,right=of bundle,minimum width=2.55cm,minimum height=1.0cm] (ar) {\textbf{AutoResearch}\\Budgeted hill climb\\Model optimization};
\node[g,right=of ar,minimum width=1.5cm,minimum height=1.0cm] (human) {Human\\acceptance};
\draw[arr] (src)--(at); \draw[arr] (at)--(bundle); \draw[arr] (bundle)--(ar); \draw[arr] (ar)--(human);
\node[s,below=9mm of bundle,minimum width=10.6cm] (runtime) {Shared agent runtime / agent job control plane \quad $\leftrightarrow$ \quad Production ML platform APIs: data, training, offline evaluation, packaging};
\draw[dasharr] (at.south) -- ++(0,-3mm) -| (runtime.north west);
\draw[dasharr] (ar.south) -- ++(0,-3mm) -| (runtime.north east);
\end{tikzpicture}
\caption{Model migration as a composition of independently usable agent loops. AutoTorch generates a training model bundle; AutoResearch can subsequently optimize the accepted model. Each loop also supports independent invocation.}
\label{fig:stack}
\end{figure*}
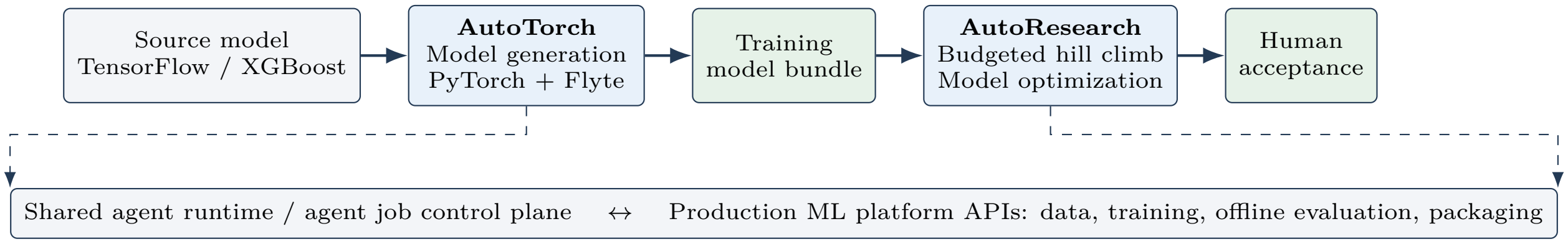

We introduce agent loops to automate this iterative process. Rather than generating model code in a single step, agents continuously generate, validate, repair, and optimize models using execution feedback. These loops operate across multiple agent sessions and GPU training jobs, with persistent state and validation mechanisms supporting long-running execution (Figure~\ref{fig:stack}). We evaluate this approach at LinkedIn across model generation, distributed training, and offline evaluation.

We implement two independently usable agent loops. \AutoTorch{} generates PyTorch models and training pipelines, including framework migration, and uses executable checks to guide repair before engineer review and full-scale training. \AutoResearch{} starts from a validated pipeline and searches for improvements through budgeted experiments. Both can run independently or be composed for migration followed by optimization.

A shared runtime supports these loops with durable execution sandboxes, persistent task state, and brokered GPU and data access. Architecture-specific skills guide candidate generation, while executable validation and experiment records provide feedback to agents. Completion remains governed by explicit checks and engineer review, not by the LLM's own assessment.

This paper describes three contributions: (1) model-generation and optimization loops with task-specific validation and search policies; (2) the shared runtime and execution controls supporting long-running tasks on production training infrastructure; and (3) deployment experience across model migration and optimization workloads at LinkedIn.

\section{Background and Problem}
\subsection{Production ML Pipelines}
Production ML pipelines combine model architectures with preprocessing, model-forward interfaces, data ingestion, distributed training configurations, and offline evaluation APIs. At LinkedIn, agents must follow these contracts to generate runnable training pipelines.

\subsection{Migration as a Use Case}
LinkedIn is migrating models from TensorFlow and XGBoost frameworks to a PyTorch-based training stack. The migration also involves framework-specific dependencies, including libraries such as TensorFlow Ranking. We use this migration to exercise the agent loops across model generation, validation, distributed training, and optimization. AutoTorch generates and validates PyTorch models and training pipelines, while AutoResearch searches for subsequent improvements. The loops can also run independently.

\subsection{Problem Formulation}
We designed the system around two common ML engineering tasks. The first is to generate a PyTorch model that works within the production training platform and reaches the required task quality. The second is to improve an existing model without disrupting its training and evaluation pipeline.

Both workflows operate within the training and offline evaluation lifecycle. API compliance includes the required preprocessing and model-forward interfaces, but does not cover online serving integration, latency or throughput SLAs, or A/B testing.

\section{System Architecture}
Each of the two agent loops consists of a set of task-specific skills and an agent harness, and can run on general-purpose coding agents such as GitHub Copilot and Claude Code. All agents run under LinkedIn enterprise agreements. Figure~\ref{fig:stack} shows the two loops composed for migration. The loops can also start independently from a task specification, an experimental training pipeline, or an accepted production model. Both run on a shared runtime (Section~\ref{sec:control}) and invoke the existing production ML platform rather than a separate agent-specific training stack.

\subsection{AutoTorch Model Generation}
\begin{figure*}[t]
\centering
\resizebox{.96\textwidth}{!}{%
\begin{tikzpicture}[node distance=3mm and 6mm]
\node[s,minimum width=1.55cm,minimum height=1.1cm] (in) {Reference TF\\or task specification\\Featurized data};
\node[b,right=of in,minimum width=2.1cm,minimum height=1.1cm] (clone) {\textbf{Stage 1}\\Clone or build\\Transfer / distill};
\node[g,right=of clone,minimum width=3.0cm,minimum height=1.1cm] (checks) {Sampled executable checks\\API + trainability: mandatory\\TF conversion: $S_{\rm TF}\geq 90$};
\node[s,right=of checks,minimum width=1.55cm,minimum height=1.1cm] (review) {Engineer\\review (PR)};
\node[b,right=of review,minimum width=2.25cm,minimum height=1.1cm] (full) {\textbf{Stage 2}\\Flyte distributed training\\Final cold start};
\node[g,right=of full,minimum width=1.55cm,minimum height=1.1cm] (out) {Checkpoint\\Offline Evals API};
\foreach \a/\b in {in/clone,clone/checks,checks/review,review/full,full/out}{\draw[arr] (\a)--(\b);}
\draw[dasharr] (checks.south) -- ++(0,-0.4) -| node[pos=.25,below,font=\scriptsize]{scores + diagnostics: repair} (clone.south);
\node[below=10mm of checks,font=\scriptsize,align=center] {Fresh generation and XGBoost replacement use code/export checks in Stage 1\\and separate post-training task-metric scoring.};
\end{tikzpicture}}
\caption{AutoTorch's two stages. Stage~1 clones a TF model or builds a new architecture, with sampled checks guiding repair. After engineer review, Stage~2 can use a warm start for a fast task-metric check before final cold-start training.}
\label{fig:stages}
\end{figure*}
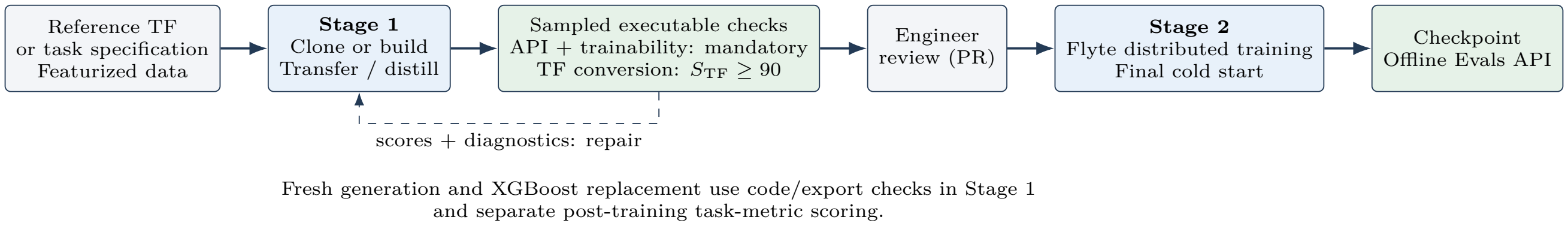

AutoTorch generates trainable PyTorch models and their training pipelines in two stages separated by engineer review (Figure~\ref{fig:stages}). Stage~1 generates the model architecture and iteratively validates and repairs it on a single GPU using sampled data. Once an engineer approves the architecture, Stage~2 builds a Flyte training pipeline and runs distributed training on multiple GPUs using production configurations and the full-scale dataset. This separation allows agents to resolve generation and validation failures before committing to full-scale training.

For migration tasks, AutoTorch takes a source model and featurized data. TensorFlow (TF) sources support weight transfer and parity checks, while XGBoost sources provide task-quality targets. For new models, AutoTorch uses a task specification and architecture-specific skills, including wide and deep models, generative recommenders, graph neural networks, TabM~\citep{gorishniy2025tabm}, and multi-task learning. The Stage~1 output is a pull request containing generated code (\code{model.py}, \code{train.py}, \code{data.py}, \code{transforms.py}, \code{eval.py}) and sampled validation results. The trained checkpoint is produced in Stage~2.

\paragraph{Stage 1: sampled generation and repair.}
For TensorFlow migration, the agent reconstructs the model in PyTorch, transfers compatible parameters using deterministic analyzers or agent-generated scripts, and applies lightweight output distillation when some parameters cannot be mapped. Each iteration evaluates the candidate on a sample of the featurized dataset. TensorFlow conversions are scored on five dimensions: output parity (35\%), task-metric parity (20\%), trainability (15\%), code style (15\%), and structural similarity (15\%). API compliance and trainability are also mandatory checks. A high cloning match requires both checks and a composite score of at least 90:
\begin{equation}\label{eq:score}
S_{\rm TF}=0.35P_{\rm io}+0.20P_{\rm task}+0.15T+0.15C+0.15H\geq 90.
\end{equation}
Here, $P_{\rm io}$ and $P_{\rm task}$ denote output and task-metric parity, while $T$, $C$, and $H$ denote trainability, code style, and structural similarity, respectively. Each component is scored from 0 to 100.

Failed checks return scores and diagnostic evidence to the agent for repair. For new models and XGBoost replacements, Stage~1 checks generated-code validity, API compliance, trainability on sampled data, and package/exportability before engineer review. Source-output parity and the TensorFlow conversion score do not apply to these models. Task-metric qualification follows full training, as described in Section~\ref{sec:scoring}.

Once the generated architecture passes its initial checks, AutoTorch opens a pull request for engineer review. Stage~2 begins after approval. Conversions scoring below 90 require explicit engineer approval and are not counted as high cloning matches.

\paragraph{Stage 2: full-scale training.}
The agent builds a Flyte-based trainer pipeline for the approved model architecture and iterates until the component runs. The pipeline is configured with the GPU count (\code{n\_gpus}) and the full-scale dataset, and packages the result as a torch package. Warm-start training from transferred or distilled parameters can provide a fast check of task-metric parity. Before downstream deployment, the model is trained from a cold start using production configurations. The resulting checkpoint is evaluated through the standard offline Evals API.

These stages measure different properties. Stage~1 evaluates conversion fidelity by comparing the clone with the TF model on identical inputs. Stage~2 verifies task quality after full-scale training, with cold-start training establishing the final PyTorch model lineage.

\subsection{AutoResearch: Hill Climb on Validated Model}
AutoResearch starts with a model that already has a working training and evaluation pipeline. The agent proposes changes to the architecture or training configuration, runs experiments, and compares the results against the current best model.

Each candidate runs in a separate Git worktree branched from the same incumbent, which remains fixed throughout the round. Experiments run in parallel, but incumbent updates and the next round wait until all trials in the current round have finished. Section~\ref{sec:orchestrate} describes the selection policy.

\paragraph{Task.} The input is a validated model plus a research plan. The model is supplied either as an existing Flyte training pipeline or as a GitHub repository with the model code. The output has three parts: a leaderboard of evaluated variants; a variant library of configurations, architectures, and code branches; and a final report of findings, dead ends, and a recommendation.

\paragraph{Research plan.} The \code{research\_plan.md} specifies the training entry point, data, primary metric, concurrency cap, and clock-hour budget. Engineers can provide free-form research directions, such as which architectures to explore, alternative losses or task-specific heads for multi-task learning, and optimizer or training configurations. They can also specify changes to avoid, such as feature crosses restricted by downstream serving requirements. When no engineer-provided research focus is available, the agent examines the model repository, training configuration, and prior experiments to develop its own research directions.

\begin{figure}[t]
\centering
\resizebox{\columnwidth}{!}{%
\begin{tikzpicture}[node distance=3.5mm]
\node[s,minimum width=6.6cm] (inc) {Validated pipeline + research plan\\Fixed incumbent throughout each round};
\node[b,below=of inc,minimum width=6.6cm] (prop) {Propose architecture / training candidates};
\node[g,below left=6mm and -3mm of prop,minimum width=1.4cm] (w1) {Worktree 1\\Train + eval};
\node[g,right=2mm of w1,minimum width=1.4cm] (w2) {Worktree 2\\Train + eval};
\node[g,right=2mm of w2,minimum width=1.4cm] (w3) {Worktree 3\\Train + eval};
\node[g,right=2mm of w3,minimum width=1.4cm] (w4) {Worktree 4\\Train + eval};
\node[s,below=13mm of prop,minimum width=6.6cm,yshift=-10mm] (sel) {Wait for all trials; select whole best improving candidate\\or keep incumbent; never merge siblings};
\node[g,below=of sel,minimum width=6.6cm] (out) {Leaderboard + variant library + report};
\draw[arr] (inc)--(prop);
\foreach \x in {w1,w2,w3,w4}{\draw[arr] (prop.south) -- (\x.north);\draw[arr] (\x.south) -- (sel.north);}
\draw[arr] (sel)--(out);
\draw[dasharr] (sel.west) -- ++(-.38,0) |- (inc.west);
\end{tikzpicture}}
\caption{AutoResearch's synchronous hill climb. A round begins from a fixed incumbent, parallel trials run in separate worktrees, and selection occurs after all results return.}
\label{fig:hill}
\end{figure}
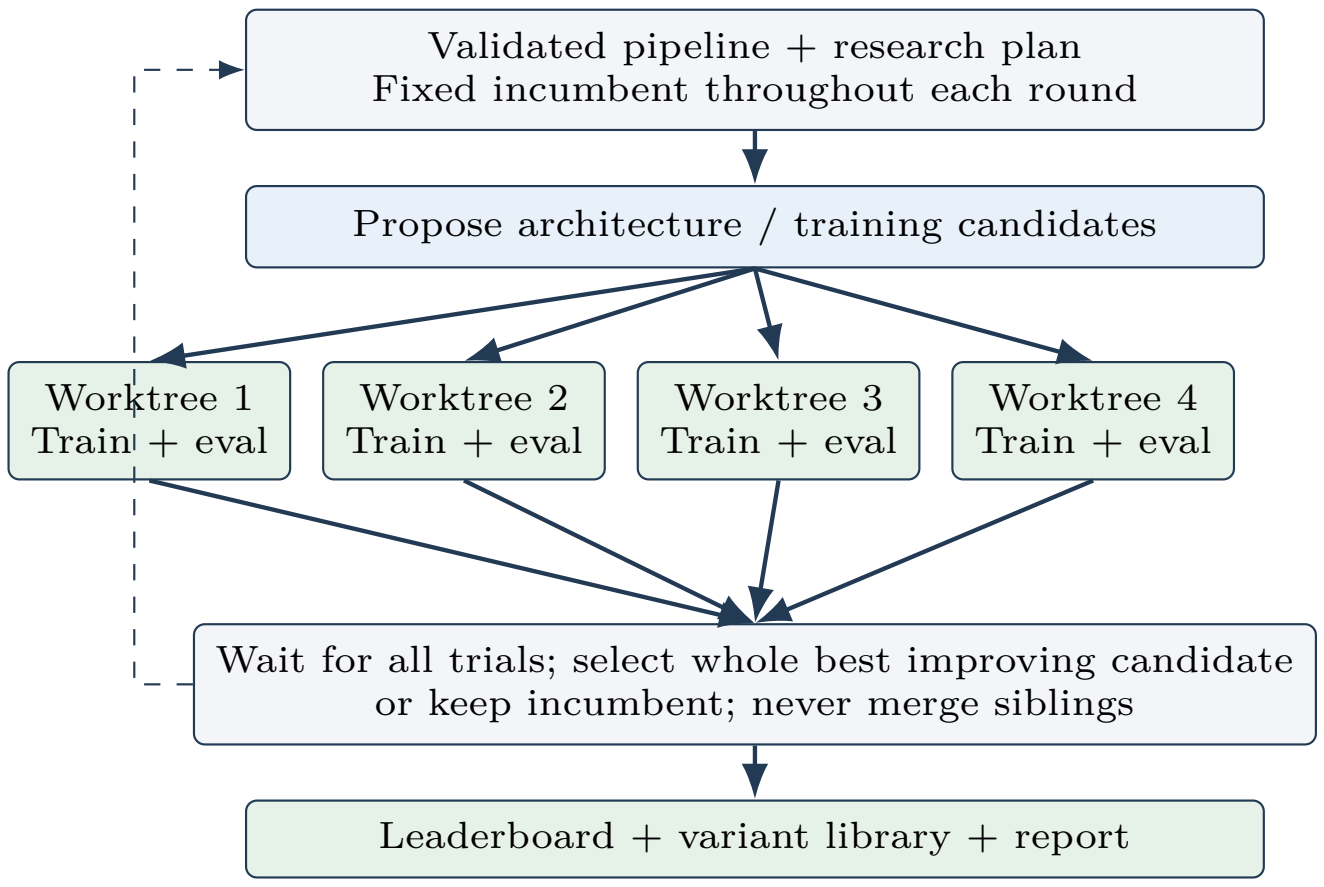

\section{Agent Control Plane}\label{sec:control}
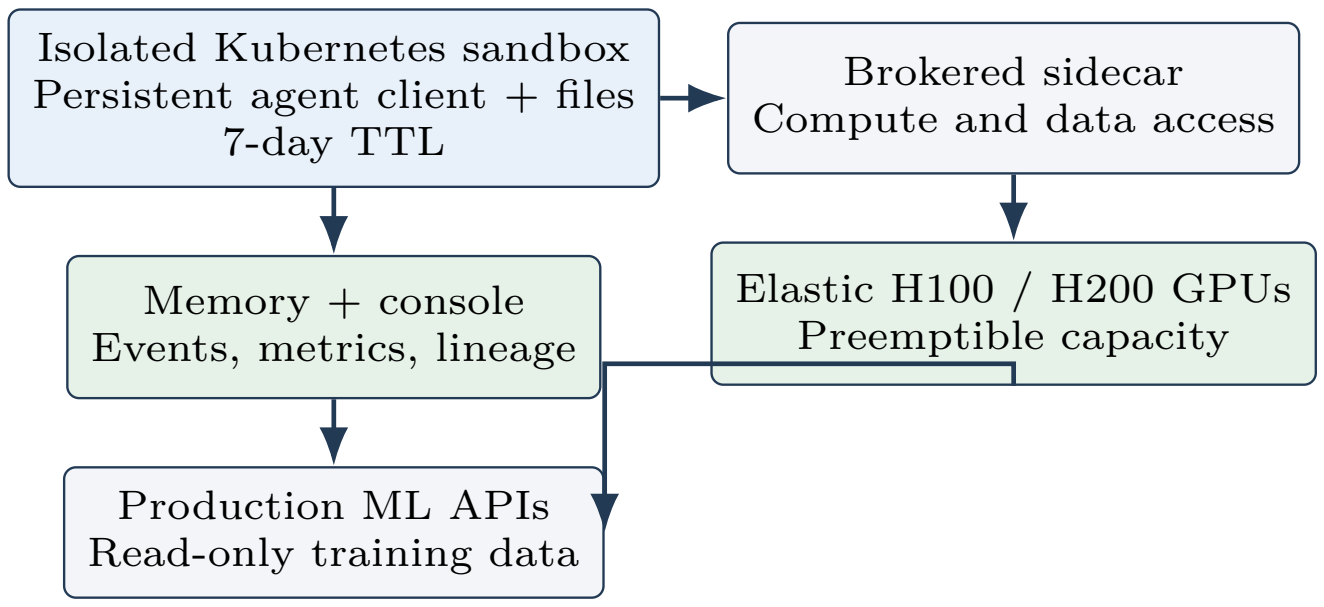
\begin{figure}[t]
\centering
\resizebox{\columnwidth}{!}{%
\begin{tikzpicture}[node distance=4mm and 4mm]
\node[b,minimum width=3.1cm,minimum height=.9cm] (agent) {Isolated Kubernetes sandbox\\Persistent agent client + files\\7-day TTL};
\node[s,right=of agent,minimum width=2.9cm,minimum height=.9cm] (broker) {Brokered sidecar\\Compute and data access};
\node[g,below=of broker,minimum width=2.9cm,minimum height=.85cm] (gpu) {Elastic H100 / H200 GPUs\\Preemptible capacity};
\node[g,below=of agent,minimum width=3.1cm,minimum height=.85cm] (mem) {Memory + console\\Events, metrics, lineage};
\draw[arr] (agent)--(broker);\draw[arr] (broker)--(gpu);\draw[arr] (agent)--(mem);
\node[s,below=of mem,minimum width=3.1cm] (api) {Production ML APIs\\Read-only training data};
\draw[arr] (gpu) -- ++(0,-.3) -| (api.east);
\draw[arr] (mem)--(api);
\end{tikzpicture}}
\caption{Durable agent sandboxes and independently allocated accelerators. Training data is available only through controlled paths and is not placed in LLM context.}
\label{fig:runtime}
\end{figure}

Coordinating agents is not the same as managing ML experiments. Delegation alone does not track which model revision produced a metric, enforce validation criteria, or preserve state across long-running GPU jobs. Our loops keep validation results, experiment history, and artifacts outside agent context, while a shared Kubernetes runtime brokers GPU resources separately (Figure~\ref{fig:runtime}).

\paragraph{Durable sandbox.} Agents execute in isolated Kubernetes pods backed by CPU or GPU resources. Sandboxes provide network storage for plans, repositories, and artifacts, with terminal session persistence and a seven-day time-to-live (TTL). GPU jobs can be preempted when production workloads need capacity. The agent client is not preempted, and its files and experiment history remain available when GPU resources return.

\paragraph{Elastic warm GPU pool.} Agents release accelerators when idle, requesting experiment pods dynamically via a broker backed by a warm H100/H200 pool. The control plane borrows idle GPU capacity from production clusters under elastic quotas, yielding capacity when production demand increases. This decouples long-running agent lifecycles from accelerator allocation.

\paragraph{Controlled data and network access.} Training jobs launched by agents read production training data through read-only, access-controlled paths; Member Data is not provided to the LLM agents' context. Operations such as trial submission and model export use brokered services; deploying a model requires engineer approval.

\paragraph{Memory.} The agentic job control plane logs all execution events and exposes a unified memory interface. A shared console tracks runs, training jobs, and metrics, maintaining artifact lineage back to originating experiments. When human intervention is required, the runtime alerts the owning team via chat while pausing the task without holding GPU resources.

\paragraph{LLM and agent configuration.} We run AutoTorch with coding and reasoning agents through Copilot. AutoTorch tasks must satisfy architecture, platform, and workflow constraints using task-specific skills, tools, and validation hooks. In AutoResearch, researcher, implementation, and evaluation roles can use different agent configurations. Experiment state, time-budget accounting, and candidate bookkeeping reside in the external orchestrator.

\section{Model Generation and Validation}\label{sec:generation}
\subsection{Model Initialization}
\paragraph{TensorFlow models.} For TensorFlow migrations, AutoTorch reconstructs preprocessing, feature ordering, embeddings, network blocks, and output heads in PyTorch. Parameter transfer uses deterministic analyzers and agent-generated scripts. The analyzers match parameters using variable names, layer types, indices, parameter roles, and tensor shapes, including layout transposition when needed. When deterministic matching fails, the agent generates custom transfer scripts to handle model-specific mappings.

When parameters cannot be mapped, AutoTorch falls back to lightweight output distillation. A structural similarity score at or below 90\% indicates a low structural match. Distillation minimizes variance-normalized mean squared error between the TensorFlow teacher and PyTorch student predictions. The teacher is evaluated without gradients, while the student's BatchNorm layers are frozen and other parameters remain trainable. By default, distillation runs for up to 500 steps with batch size 32, stopping early when the loss falls below 0.001. The default conversion path uses synthetic inputs.

\paragraph{No transferable state.} For non-cloneable models (e.g., XGBoost or custom frameworks), AutoTorch constructs a PyTorch replacement and trains it directly on labeled data using the source model solely as a quality benchmark.

\paragraph{New models.} For new model generation, AutoTorch uses task specifications, feature schemas, target metrics, and permitted architecture families to build and train candidate models using standard validation checks.

\subsection{Platform API Compliance}
Candidates must utilize supported platform APIs for data loading, training, and evaluation. We verify compliance by executing generated code against these interfaces, returning failures for repair rather than relying on static inspection alone.

\subsection{Trainability Probe}
The trainability probe performs a lightweight single-GPU optimization run to verify loss reduction and detect numerical issues, nonfinite gradients, or structural incompatibilities before allocating distributed training resources. By default, this consists of 30 training steps with batch size 256. The probe checks whether gradients flow, losses decrease, and model outputs remain numerically stable.

It also detects saturated or nearly constant outputs and unusually low initial losses. A decreasing loss trend with positive improvement passes the loss-reduction check. A 5\% relative decrease earns full credit for this component; smaller positive decreases can still pass.

\subsection{Full-Scale Distributed Training}
After engineer review, candidates advance to full-scale distributed training on multiple GPUs using production configurations. For TensorFlow migrations, a warm start from transferred or distilled parameters can verify task-metric parity through faster convergence. This offers an early indication that the PyTorch model can recover the reference model's quality.

Before downstream deployment, models are trained from a cold start in the PyTorch pipeline to establish reproducible training lineage independent of transferred TensorFlow weights. Models with string features require cold starts because LinkedIn's TensorFlow and PyTorch training stacks use different hashing and tokenization transforms, making direct reuse of feature-dependent weights unreliable (Figure~\ref{fig:stages}).

\subsection{Output Predictions Validation}
For high-fidelity model cloning, candidates must accurately reproduce the source model's prediction outputs on the same sampled inputs. The representation evaluated depends on task semantics: logits are compared for binary classification heads, softmax probabilities for multi-class heads, and raw prediction scores for ranking models. Evaluating prediction values rather than discrete class labels catches differences that label agreement alone would miss.

Output parity ($P_{\rm io}$) compares predictions on identical inputs. For scalar outputs, output parity combines correlation (40\%), tolerance-based agreement (40\%; absolute tolerance $10^{-3}$, relative tolerance 1--5\%), and near-exact match (20\%; absolute difference at most $10^{-3}$). Near-exact matching does not require bitwise equality. For vector outputs, cosine similarity and near-exact match contribute equally (50\% each).

Stage~1 uses randomly sampled inputs for lightweight distillation and conversion scoring, without enforcing disjointness between the two samples. The resulting parity scores guide conversion and iterative repair rather than serving as independent estimates of model equivalence. Final task quality is evaluated separately after full-scale training using the standard offline evaluation pipeline.

\subsection{Primary-Metric Validation}
Task-metric parity ($P_{\rm task}$) compares the primary offline metric of a TensorFlow reference and its PyTorch candidate on the same data. For AUC, the score is zero if either model performs at or below chance (0.5). AUC values between 0.5 and 0.55 reduce the parity score to account for weak predictive signal. Once both AUCs reach 0.55, the parity score depends only on their closeness: each absolute AUC gap of 0.01 reduces the score by two points from 100, with a floor of zero.

The same principle applies to other task metrics, including NDCG@K, Precision@K, and Recall@K for ranking, or RMSE for regression. Source and candidate use the same offline data and metric definition, including ranking cutoff K. Metric direction and no-signal baselines are task-specific; the AUC thresholds do not apply to these other metrics. Stage~1 uses task-metric parity to guide conversion repair. Final task quality is evaluated separately after full-scale training. For newly generated models without a source, the task specification provides the quality target.

\subsection{Multi-Dimensional Scoring}\label{sec:scoring}
TensorFlow conversions use the composite score and mandatory checks defined in Equation~\ref{eq:score}. A conversion qualifies as a high cloning match when its score reaches 90 and the mandatory API and trainability checks pass. Lower-scoring conversions require explicit engineer approval before Stage~2. The weights and threshold are fixed in this deployment; we have not evaluated their sensitivity or calibration against full-training outcomes.

For new models, there is no reference implementation to compare against. We instead score trainability (35\%), metric achievement (45\%), and code quality (20\%) after training. Before full-scale training, the generated code must pass validity and package/export checks. Post-training qualification requires the generated model to be trainable, pass API compliance checks, and meet the task-metric target supplied in its specification. The target is a minimum for higher-is-better metrics and a maximum for lower-is-better metrics; no task-independent metric cutoff is used.

\subsection{Model Packaging and Handoff}
AutoTorch outputs a trainable model bundle containing implementation code, configurations, evaluation adapters, and export metadata for downstream productionization or review.

\begin{figure}[t]
\centering
\resizebox{\columnwidth}{!}{%
\begin{tikzpicture}[node distance=4mm]
\node[b,minimum width=5.4cm] (skills) {Architecture skills: GNN / TabM / MTL\\Platform skills: dataloader / trainer / Evals};
\node[s,below=of skills,minimum width=5.4cm] (bundle) {Agent builds trainable PyTorch bundle};
\node[g,below=of bundle,minimum width=5.4cm] (checks) {Executable hooks: API + trainability required\\TF: parity + style + structure; $S_{\rm TF}\geq 90$};
\node[b,below=of checks,minimum width=5.4cm] (full) {Engineer review $\rightarrow$ full-scale training\\Final task-metric evaluation};
\draw[arr] (skills)--(bundle);\draw[arr] (bundle)--(checks);\draw[arr] (checks)--(full);
\draw[dasharr] (checks.west) -- ++(-.4,0) |- (bundle.west);
\end{tikzpicture}}
\caption{Skills propose architectures, and executable checks return failure evidence for repair. New-model generation uses separate scoring after full training.}
\label{fig:validation}
\end{figure}
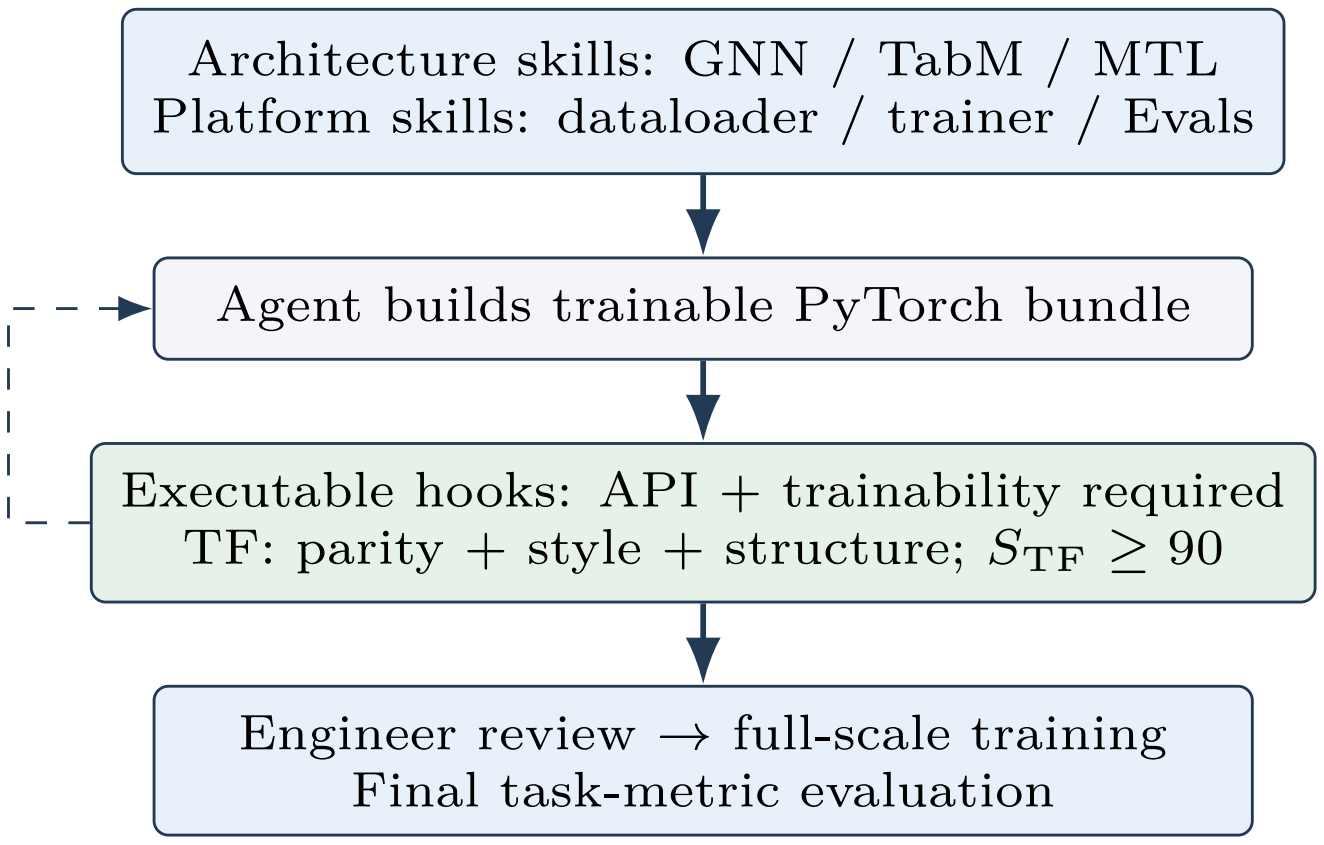

\section{Budgeted Model Research}\label{sec:research}
\subsection{Search Objective and Constraints}
AutoResearch searches for model improvements within a fixed wall-clock time budget. Starting from a validated baseline $c_0$, it evaluates candidate changes using the model's existing training and evaluation pipeline. The current best model, or incumbent, is updated only after a round completes. The search returns this incumbent, or the original model if no round improves it.

\subsection{Research Direction Search}
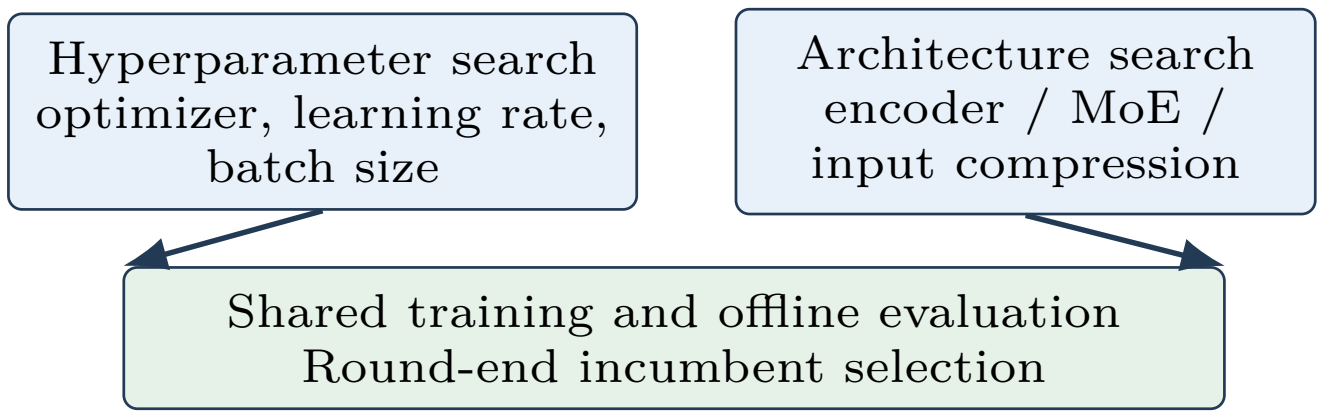
\begin{figure}[t]
\centering
\resizebox{\columnwidth}{!}{%
\begin{tikzpicture}[node distance=5mm and 5mm]
\node[b,minimum width=2.8cm] (hp) {Hyperparameter search\\optimizer, learning rate,\\batch size};
\node[b,right=of hp,minimum width=3cm] (arch) {Architecture search\\encoder / MoE /\\input compression};
\node[g,below=8mm of $(hp)!0.5!(arch)$,minimum width=5.7cm] (ev) {Shared training and offline evaluation\\Round-end incumbent selection};
\draw[arr] (hp.south)--(ev.north west);\draw[arr] (arch.south)--(ev.north east);
\end{tikzpicture}}
\caption{Hyperparameter and architecture search share the same evaluation loop and round-end selection policy.}
\label{fig:modes}
\end{figure}

AutoResearch proposes model changes based on the existing code, configurations, and previous experiments (Figure~\ref{fig:modes}). Engineers can also specify research directions. Each proposal identifies the components to change, the expected effect, the target metric, and a minimal experiment. When improvements plateau, a dedicated skill explores alternative directions. Search stops when the clock-hour budget or configured stopping criterion is reached.

\subsection{Experiment Orchestration}\label{sec:orchestrate}
We use an external Python orchestrator to manage the search. It tracks experiment history, submits parallel training jobs, records metrics, and maintains the current incumbent.

\begin{figure}[t]
\centering
\resizebox{\columnwidth}{!}{%
\begin{tikzpicture}[node distance=3mm and 3mm]
\node[b,minimum width=1.8cm] (r1) {Round 1\\Propose};
\node[b,right=of r1,minimum width=1.8cm] (r2) {Round 2\\Propose};
\node[b,right=of r2,minimum width=1.8cm] (rr) {Round R\\Propose};
\node[g,below=of r1,minimum width=1.8cm] (e1) {$\leq N$ trials\\in parallel};
\node[g,below=of r2,minimum width=1.8cm] (e2) {$\leq N$ trials\\in parallel};
\node[g,below=of rr,minimum width=1.8cm] (er) {$\leq N$ trials\\in parallel};
\node[s,below=of e1,minimum width=1.8cm] (s1) {Wait all\\select};
\node[s,below=of e2,minimum width=1.8cm] (s2) {Wait all\\select};
\node[s,below=of er,minimum width=1.8cm] (sr) {Wait all\\select};
\foreach \a/\b in {r1/e1,r2/e2,rr/er,e1/s1,e2/s2,er/sr}{\draw[arr] (\a)--(\b);}
\draw[arr] (s1.east)--(r2.west);\draw[arr] (s2.east)--(rr.west);
\node[below=5mm of s2,font=\scriptsize] {Shared elapsed budget $B$; return incumbent at timeout};
\end{tikzpicture}}
\caption{Synchronous AutoResearch rounds. Each round waits for all trials, then retains or replaces the incumbent. All rounds share one elapsed-time budget.}
\label{fig:rounds}
\end{figure}
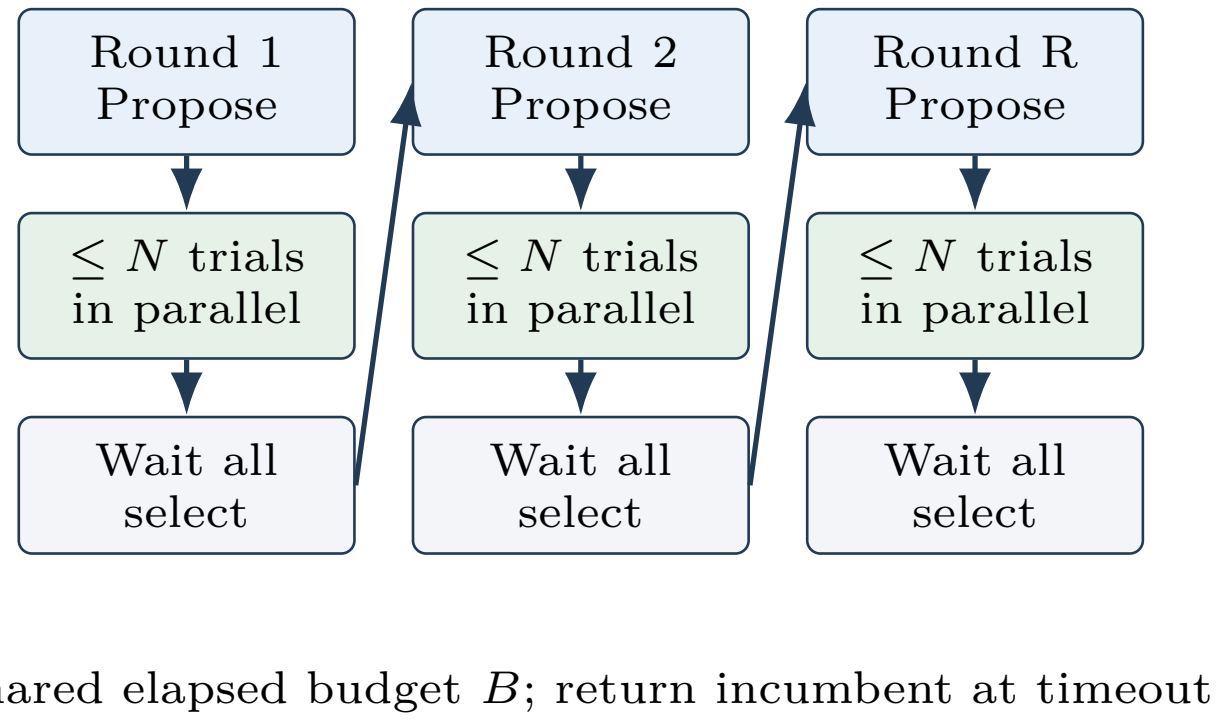

Each search has a wall-clock time budget and a configurable concurrency limit, typically four experiments. Results are recorded as trials finish, but no replacement trial is launched and the incumbent is not updated until the round completes. The orchestrator then selects the best candidate that improves the incumbent; otherwise, it keeps the incumbent. Candidates are selected as complete implementations, without automatically merging changes from sibling worktrees. Completed-round results guide the next proposals (Figure~\ref{fig:rounds}).

The time budget includes waiting for slower trials at the round barrier. This policy trades immediate reuse of idle slots for comparison against a fixed incumbent. Experiment history persists across agent sessions.

\subsection{Candidate Evaluation}
During search, candidates are ranked against the incumbent using the pipeline's primary metric. After search, the selected candidate is compared with the original baseline on held-out data reserved for final evaluation~\citep{cawley2010overfitting}. The held-out results do not guide proposals, incumbent updates, or selection of another candidate. The improvement results in Table~\ref{tab:results} use this final held-out comparison.

\subsection{Self-Repairing Harness}
Not every failed experiment indicates a bad model. Some failures come from the training environment, dependencies, or the harness itself.

When this happens, the agent uses the error logs and execution context to diagnose the issue and propose a fix. Harness changes are submitted as pull requests and go through normal code review and integration tests. Successful repairs can also inform future runs through persistent memory, reducing the need to diagnose the same issue repeatedly.

\begin{table*}[t]
\centering\scriptsize
\caption{Production outcomes and runtime. AutoTorch latencies refer to the indicated intervals; end-to-end time includes engineer review. The two denominator cohorts differ.}
\label{tab:results}
\begin{tabularx}{\textwidth}{@{}p{.085\textwidth}p{.31\textwidth}p{.18\textwidth}X@{}}
\toprule
\multicolumn{2}{@{}l}{\textbf{1. Deployment outcomes}} & \textbf{Result} & \textbf{Scope} \\
\midrule
1.1 & AutoTorch generation runs & 115 & Workflow executions; not necessarily distinct models \\
1.2 & Operational completion & 92 of 115 (80\%) & Generation and full training; denominator: all attempted runs \\
1.3 & Post-training quality pass & 77 of 92 (83.7\%) & Evaluated model artifacts from completed runs; TF clones also require a conversion score $\geq90$ \\
1.4 & Offline primary-metric difference & Typically within 1--5\% & Relative to reference after full training; initialization modes not disaggregated; not a deployment gate \\
1.5 & AutoResearch improvements & 28 of 100 (28\%) & Improved original baseline on final held-out evaluation \\
\midrule
\multicolumn{2}{@{}l}{\textbf{2. Measured AutoTorch execution}} & \textbf{p50 / p90} & \textbf{Range} \\
\multicolumn{4}{@{}l}{\emph{Stage 1: sampled generation and repair}}\\
2.1 & Generate--evaluate--repair iterations & 4 / 10 & 1--17 \\
2.2 & Stage 1 wall time & 6.3 h / 32 h & 0.6--46 h \\
\multicolumn{4}{@{}l}{\emph{Stage 2: distributed training, evaluation, and export}}\\
2.3 & Successful training job & 33 min / 2.5 h & 2 min--6.4 h \\
2.4 & Train $\rightarrow$ evaluate $\rightarrow$ bundle $\rightarrow$ export & 2.8 h / 3.8 h & 1.0--5.3 h \\
\multicolumn{4}{@{}l}{\emph{End-to-end: generation start, including engineer review}}\\
2.5 & First trained model & 2.3 d / 3.7 d & 0.1--6.1 d \\
2.6 & Exported model & 2.4 d / 5.8 d & 0.1--9.2 d \\
\midrule
\multicolumn{2}{@{}l}{\textbf{3. Research and migration time}} & \textbf{Result} & \textbf{Scope} \\
3.1 & AutoResearch search runtime & 48--72 h & Reported wall-clock range, not percentiles \\
3.2 & AutoResearch optimization portion & About 3 days & Estimated component of an agent-assisted migration project \\
3.3 & Manual migration & About 8 weeks & Team-estimated duration to a pipeline of comparable offline quality \\
3.4 & Agent-assisted migration & About 2 weeks & Team-estimated duration to the same endpoint \\
3.5 & End-to-end migration acceleration & Approximately $4\times$ & Ratio of project-duration estimates; not engineer-hours \\
3.6 & Model-specific manual fixes & About 2 engineer-days, when needed & Conditional human effort; not elapsed project time \\
\bottomrule
\end{tabularx}
\vspace{2pt}\parbox{\textwidth}{\footnotesize\emph{Timing note.} Recorded stage latencies and end-to-end latencies refer to distinct intervals and applicable run subsets, and their percentiles are not additive. Migration-project durations and component estimates are team reported; engineer-days measure labor, not elapsed time.}
\end{table*}

\section{Evaluation Methodology}\label{sec:evaluation}
\paragraph{Research questions.} We organize deployment observations around three questions: (RQ1) Can agent loops generate, train, and evaluate models on production ML infrastructure? (RQ2) How do external control mechanisms enforce execution and completion constraints? (RQ3) Can agent loops like AutoResearch improve model quality under fixed clock-hour budgets?

\subsection{RQ1: ML Framework Migration}
Since May 2026, we have used AutoTorch for TensorFlow migration and XGBoost replacement. Operational completion requires generation and full-scale training. Quality qualification checks task quality and applicable cloning criteria only for completed runs. Table~\ref{tab:results} reports completion over all attempts and quality passing over completions; neither rate measures engineer overrides. \textbf{Engineers can reuse code from incomplete runs for debugging and further iteration.}

Table~\ref{tab:cases} presents five AutoTorch cases spanning greenfield migration, TensorFlow-to-PyTorch porting, and tree-to-neural replacement. The agents delivered PyTorch models and training pipelines, using validation feedback to repair data processing, evaluation, and architecture mismatches.

\paragraph{Execution latency.} Stage~1 generation and repair took 6.3 hours at p50 and 32 hours at p90. Among exported models, end-to-end generation to export, including engineer review, took 2.4 days at p50 and 5.8 days at p90 (Table~\ref{tab:results}).

\paragraph{Migration time.} Engineering teams report that migrating a pipeline with comparable offline primary-metric quality typically takes eight weeks manually, compared with two weeks using the agent stack---a $4\times$ speedup in end-to-end migration. The comparison measures elapsed project duration, not active engineer-hours. Measured AutoTorch export time and estimated full-project migration duration cover different workflow boundaries.

\begin{table*}[t]
\centering\scriptsize
\caption{Selected agent-loop cases (AT: AutoTorch; AR: AutoResearch). These examples are not randomly sampled.}
\label{tab:cases}
\begin{tabularx}{\textwidth}{@{}p{.06\textwidth}p{.22\textwidth}p{.43\textwidth}X@{}}
\toprule
\textbf{Case} & \textbf{Model / use case} & \textbf{Agent action} & \textbf{Primary task metric}\\
\midrule
\multicolumn{4}{l}{\textbf{AutoTorch --- generation and migration}}\\
AT-1 & Two-tower video retrieval & Delivered PyTorch model and training pipeline based on source model & Recall@200 matched source model\\
AT-2 & XGBoost $\rightarrow$ TabM; job-posting classification & Changed loss to per-example binary cross-entropy & AUC matched\\
AT-3 & TensorFlow Ranking-based CTR model & Ported to PyTorch with weight transfer; fixed data and evaluation mismatches & AUC matched\\
AT-4 & TensorFlow multi-task model (200M embedding parameters) & Ported to PyTorch; repaired preprocessing, sparse gradients, and task-output handling & Embedding outputs matched (output-parity check)\\
AT-5 & TensorFlow graph neural network & Ported to a PyTorch GNN; restored node encoder and concatenation combiner & Downstream classifier AUC matched\\
\midrule
\multicolumn{4}{l}{\textbf{AutoResearch --- model optimization}}\\
AR-1 & GNN connection recommendation & Reduced encoder to one wider layer; increased neighbor sampling; averaged four inference passes; tuned training & Recall@1000 $+9.3\%$ vs. baseline\\
AR-2 & MLP classifier & Widened hidden layers $[24,12]\rightarrow[64,32]$ & Held-out primary AUC $+1.20\%$ vs. baseline\\
AR-3 & SLM (0.6B) + MLP classification & Compared linear, MMoE, PLE, and DCNv2 heads with engineer guidance & $p_{\rm click}$ AUC $+1.2\%$ vs. MLP baseline (validation)\\
\bottomrule
\end{tabularx}
\vspace{2pt}\parbox{\textwidth}{\footnotesize\emph{Notes.} Matching does not imply bitwise equivalence. AT-4 reports embedding-output parity, not a primary task metric. AR-1 combines several changes on a fixed evaluation fixture; independent held-out status is not reported. AR-3 is a task-specific validation result, not an additional held-out win in Table~\ref{tab:results}.}
\end{table*}

\subsection{RQ2: Harness Design for External Control under Constraints}
\paragraph{Iteration overhead.} Setup and the first scored iteration took 2.5 hours at p50 and 15 hours at p90, compared with 46 minutes and 3.2 hours for subsequent repair iterations. The first interval includes model construction and initial validation, whereas later iterations reuse the initialized workflow.

\begin{figure*}[t]
\centering
\resizebox{.96\textwidth}{!}{%
\begin{tikzpicture}[node distance=4mm and 4mm]
\node[s] (agent) {Coding agent\\LLM + skills};
\node[b,right=of agent,minimum width=3.4cm] (hooks) {\textbf{AutoTorch: executable hooks}\\Architecture, numerical, API checks\\Pass / failure evidence};
\node[g,right=of hooks] (review) {Eligible for\\engineer review};
\draw[arr] (agent)--(hooks); \draw[arr] (hooks)--(review);
\draw[dasharr] (hooks.south) -- ++(0,-.65) -| node[pos=.3,below,font=\scriptsize] {failure $\rightarrow$ repair} (agent.south);
\node[s,right=1.3cm of review] (orchestrator) {\textbf{AutoResearch: orchestrator}\\Candidate ledger, history, budget\\Round-end selection};
\node[b,right=of orchestrator] (sessions) {Agent sessions\\Fresh reasoning context};
\node[g,below=of sessions] (jobs) {Brokered GPU jobs\\Persistent metrics};
\draw[arr] (orchestrator)--(sessions); \draw[arr] (sessions)--(jobs); \draw[arr] (jobs.west) -| (orchestrator.south);
\end{tikzpicture}}
\caption{Harness-specific execution control. AutoTorch uses hooks to gate each generate--evaluate--repair iteration. AutoResearch maintains persistent state and round-end selection externally to agent sessions.}
\label{fig:harness}
\end{figure*}
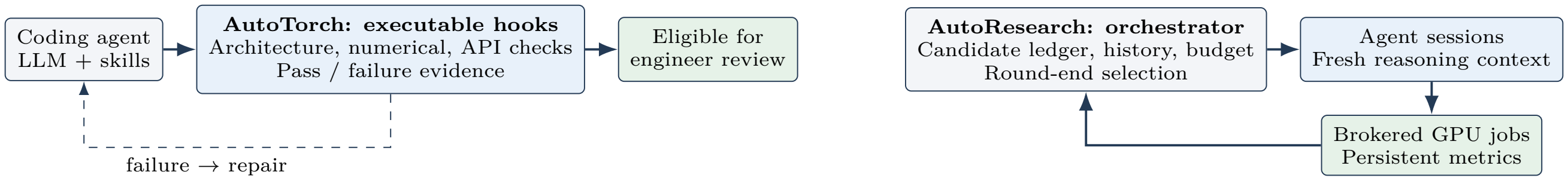

AutoTorch uses executable hooks to enforce business and platform rules, including API compliance, trainability, and conversion fidelity. Failed checks return diagnostic feedback for the next repair iteration. AutoResearch instead maintains experiment history, candidate selection, and time budgets in an external Python orchestrator. This keeps execution state separate from agent reasoning across long-running experiments and multiple agent sessions (Figure~\ref{fig:harness}).

\subsection{RQ3: Agent-Guided Model Optimization}
AutoResearch improved the primary metric on final held-out evaluation in 28 of 100 targeted pipelines (Table~\ref{tab:results}). If no candidate improved the search-validation metric, the incumbent was retained; held-out results were reported separately.

AutoResearch supports two modes of experimentation: expert-steered and autonomous. Domain experts can guide architectures, training configurations, and constraints to explore; without that guidance, the agent uses the model repository, configurations, and previous results to propose experiments. Both modes use the same iterative search loop.

Table~\ref{tab:cases} shows three AutoResearch cases. For a GNN recommender, AutoResearch explored 49 trials and identified a configuration that improved Recall@1000 by 9.3\%. Widening an MLP classifier improved held-out AUC by 1.2\%. For an SLM (0.6B) engagement model, an engineer used AutoResearch to compare prediction-head architectures, including MMoE~\citep{ma2018mmoe}, PLE~\citep{tang2020ple}, and DCNv2~\citep{wang2021dcnv2}, and reported obtaining decisive task-level results in days rather than weeks of manual exploration. DCNv2 improved two engagement metrics while meeting the relevance guardrail. These examples show how domain-expert guidance and agent-generated proposals can accelerate model experimentation.

\section{Related Work}
\paragraph{Model migration.} Nikolov et al.~\citep{nikolov2026migration} use playbooks and specialized agents for TensorFlow-to-JAX migration, evaluating complex internal migrations with checklist-based LLM judges. Liang et al.~\citep{liang2026migration} combine in-context examples with source-execution oracles and iterative debugging for PyTorch-to-JAX migration. Our workflow combines TensorFlow-to-PyTorch parameter transfer and execution-based qualification with full-scale training and offline evaluation.

\paragraph{ML experimentation agents.} AIDE explores code search for model implementation~\citep{jiang2025aide}, MLE-STAR refines pipelines using retrieved knowledge~\citep{nam2025mlestar}, and Auto-RecSys combines asynchronous recommendation experiments with persistent memory and a separation between reasoning guidance and deterministic execution~\citep{li2026autorecsys}. Our system combines model generation and optimization with production training and offline evaluation in a shared platform.

\paragraph{Autonomous research.} Karpathy's autoresearch uses markdown instructions to guide single-GPU training experiments~\citep{karpathy2026autoresearch}. AutoResearch was developed independently and does not use code from autoresearch. R\&D-Agent separates research planning from implementation and evaluation~\citep{yang2025rndagent}, while The AI Scientist automates idea generation and experimental execution~\citep{lu2024scientist}. AutoResearch applies related propose--implement--evaluate loops to existing production training pipelines, with synchronous parallel trials and a shared elapsed-time budget.

\paragraph{Agent runtimes.} AgentScope provides multi-agent communication and fault tolerance~\citep{gao2024agentscope}. The agentic job control plane complements this by providing isolated sandboxes and brokered access to compute, production data, and ML platform APIs for long-running workflows.

\paragraph{Program search and optimization.} AlphaEvolve couples code changes to automated evaluation~\citep{novikov2025alphaevolve}. Random search~\citep{bergstra2012random}, Optuna~\citep{akiba2019optuna}, and Vizier~\citep{golovin2017vizier} provide fixed-space optimization baselines. Our evaluation uses live production workloads rather than public benchmarks~\citep{huang2024mlagentbench,chan2025mlebench}.

\section{Discussion and Limitations}
The agent loops automate substantial parts of model generation and optimization, but engineers remain responsible for reviewing generated code, handling model-specific requirements, and making deployment decisions. Executable validation helps identify numerical and platform-integration failures, but it cannot establish model equivalence across all inputs.

A central limitation is potential skew between offline and online task metrics. Models that match or improve offline metrics can behave differently under production traffic, shifting data distributions, and user feedback. Our evaluation covers training and offline model quality; online serving integration and A/B testing remain outside the scope of the agent stack.

The system also depends on LinkedIn's existing preprocessing, training, and evaluation interfaces. Adapting these agent loops to another ML platform would require integrating with its corresponding interfaces.

\section{Conclusion}
We presented AutoTorch and AutoResearch, two agent loops for model generation, framework migration, and optimization within LinkedIn's PyTorch training platform. AutoTorch uses executable validation to guide model generation and repair, while AutoResearch keeps experiment state across long-running optimization tasks. Engineering teams report a $4\times$ reduction in elapsed migration time while recovering comparable offline model quality. The stack automates much of this work while retaining engineer review and established production training controls.

\section*{AI Usage Disclosure}
We used Copilot with OpenAI and Anthropic models and Gemini to assist with manuscript editing, literature search, and the preparation of LaTeX and TikZ figures. The authors reviewed and revised the generated content. The reported system implementation and experimental results were provided by the authors and were not generated by these AI tools. The authors take responsibility for the accuracy and integrity of the final manuscript.
\end{document}